# The physics of osmotic pressure

M. G. Bowler

*Department of Physics, University of Oxford, Keble Road, Oxford OX1 3RH* [a)]

**Abstract**

The proper application of the virial theorem to the origin of osmotic pressure provides a simple, vivid and complete explanation of the physical processes involved; osmosis is driven by differential solvent pressures. This simple and intuitive notion is widely disregarded, but the virial theorem allows an exact formulation and unifies a number of different treatments. It is closely related to a kinetic treatment devised by Ehrenfest over 100 years ago.

## I. INTRODUCTION

Consider two reservoirs, one filled with a solvent such as pure water and the other with a solution, such as salt or sugar in water (Fig.1). If the two are separated only by a semi-permeable membrane that passes water in both

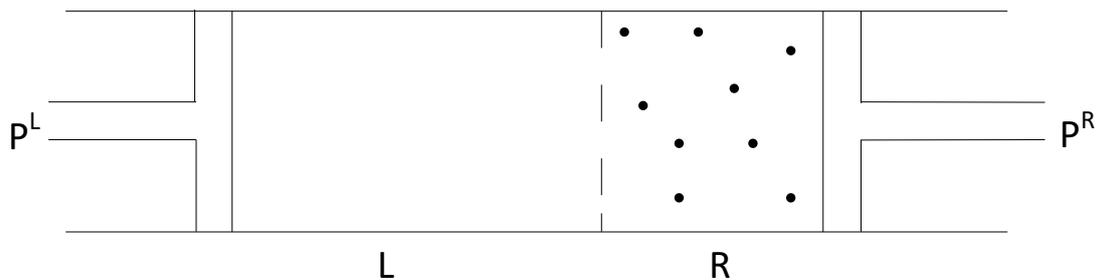

*Figure 1 shows two compartments separated by a semi-permeable membrane. The left contains pure solvent (water) and the right contains solution (such as salt in water). Water streams from left to right until the right hand pressure exceeds the left hand pressure by the osmotic pressure.*

directions but does not pass salt at all, then water streams into the solution and the flow is only arrested if the external pressure applied to the solution exceeds that applied to the solvent by the *osmotic pressure* $\Pi$. Further increase of the solution pressure ($p^R$ in Fig.1) drives water back into the left hand compartment. The osmotic pressure is given to a good approximation by

$$\Pi = n_s kT \qquad (1)$$

Eq. (1) is known as van't Hoff's law and looks like an ideal gas relation – in (1) $T$ is the temperature, $k$ is Boltzmann's constant and $n_s$ is the number density of solute units (atoms, ions or molecules).

In the thermodynamics of an ideal solution, entropy of mixing affects the chemical potential of the water component. Equating the chemical potentials of water on each side of the semi-permeable membrane then yields (1) for equilibrium (see for example [1]). However, the strength of thermodynamics is that it yields results without an understanding of underlying mechanisms, but this is its weakness. It seems that there is yet considerable confusion (and misinformation) about the origin of the forces that drive solvent from the water side into the solution and the role of the semi-permeable membrane, see for example, [2] more recently. [3] The forces can seem enormous (osmotic pressures of 10s of atmospheres for dilute solutions) and their relation to entropy of mixing (which does not involve any changes of internal energy, temperature or pressure) mysterious.

That hydrostatic pressure differences are involved suggests a mechanism directly related to everyday experience. If two containers are linked at a level corresponding to different water pressures, then water flows until the pressures equalize. Then in Fig.1 it might be the water pressure in the left hand compartment that drives water to the right against a lower water pressure. After all, the solute will exert a pressure and if initially the left and right hand pressures are equal, surely that must mean a lower water pressure in solution? This mechanism is intuitively appealing and the proposal ancient, but it has never previously been adequately formulated.

There is in classical mechanics the *virial theorem* that relates the motions of atoms and interactions between them to fluid pressure exerted on the walls of a container. The proper application of this theorem to the atoms of solvent and solute in solution formulates quantitatively and correctly the notion of hydrostatic water pressure differences driving osmosis. The simple analysis presented below answers completely and unambiguously the questions of how the forces arise and the role of the membrane, revealing explicitly the parts played by molecular bombardment and by forces acting within pairs of water molecules, pairs of solute molecules and in particular between water and solute molecules. It also casts its own light on the 'questions about osmosis' posed in Ref.3.

The virial theorem was first applied in this century to the problem of osmotic pressure, as far as I know, by F G Borg.[4] Unfortunately certain steps in his treatment are confused and, I believe, erroneous. It was applied in a rather different way in 1915 by Ehrenfest[5]; his work is neither confused nor erroneous, but he claimed only that it applies to dilute solutions. It can be made more general and then becomes equivalent to the new analysis presented here.

## II. THE VIRIAL THEOREM AND INTERMOLECULAR FORCES

The origin of osmotic pressure for the particular case of a mixture of (approximately) ideal gases is very straightforward. If the gaseous solution and the solvent are at the same pressure, the solvent gas is at a lower partial pressure on the solution side. The (partial) pressure difference drives the solvent gas through the membrane and the role of the membrane is to contain the solute gas. For an ideal gas interactions between all molecules are supposed negligible, other than for establishing thermal equilibrium. This case should be contained in a treatment valid for liquid solutions, more complicated because in liquids inter-molecular forces are of great importance. For such solutions, formal partial pressures do not exist, yet solvent and solute can be assigned well-defined pressures in certain circumstances. This is where the virial theorem comes in.

The virial theorem in the classical mechanics of particles[6] can relate fluid pressure exerted on the walls of a container of volume $V$ to the kinetic and potential energies of the particles confined within it. A proof in suitable form is given in Appendix A of Ref.4; another suitable treatment may be found in Ref.7. For fluid pressure the virial theorem can be written

$$3pV = 2\left\langle \sum_i T_i \right\rangle + \left\langle \sum_i r_i \cdot \sum_j F_{ij} \right\rangle \qquad (2)$$

On the right hand side the brackets denote time averaging. The first term on the right is the sum over all particles of kinetic energy and the second term is the time average of the sum over all particles of the scalar product of the position vector of particle *i* with the force acting upon it due to all other particles. It is important to note that an individual term $r_i . F_{ij}$ in the double sum can have no physical significance because it is not invariant under a shift in the origin of the coordinates. However, because the forces are all between pairs of particles *i*, *j* the terms in the double sum couple up so that the last term in (2) can be written

$$\left\langle \sum_{pairs} F_{ij} .(r_i - r_j) \right\rangle$$

The individual terms in this sum over pairs (Fig.2) are translation invariant (because a shift in the origin of coordinates vanishes in the difference between the two position vectors) and hence can have physical significance.

For a system in thermal equilibrium and with negligible interatomic forces, (2) yields the equation of state of an ideal gas. More generally, (2) can be written

$$p = \tfrac{2}{3}\tfrac{1}{V}\left\langle \sum_i T_i \right\rangle + \tfrac{1}{3}\tfrac{1}{V}\left\langle \sum_{pairs} F_{ij}.(r_i - r_j) \right\rangle \qquad (3)$$

where the first term is the contribution of the kinetic energy to the pressure (kinetic pressure *K*) and the second the contribution of the interatomic forces, related to a potential energy density, *P*. For a gas the interatomic term *P* is negligible. For a liquid both *K* and *P* are very much larger than for a dilute gas, of comparable magnitude and opposite sign. The quantity *P* is dominated by cohesive forces and so is a negative pressure, tension.

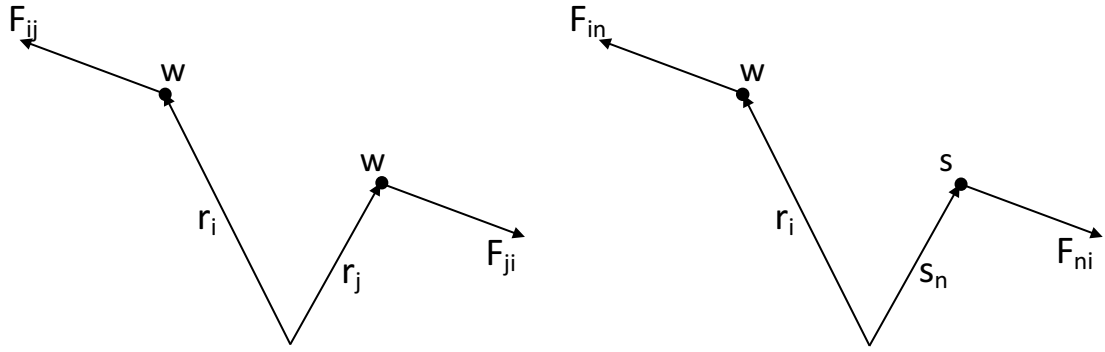

*Figure 2 The equal and opposite forces between two molecules are shown (drawn as repulsive for clarity) for two water molecules ww in the left hand panel and between water and solute ws in the right hand panel. The w and s coordinate vectors are drawn from some arbitrary origin.*

### III. APPLICATION TO OSMOSIS

Considering two compartments as in Fig.1, but isolated; (3) can be applied separately to each. The pressure $p^R$ exerted by the solution in the right hand compartment is usefully split into five components

$$p^R = K_w^R + P_{ww}^R + P_{ws}^R + K_s^R + P_{ss}^R \qquad (4)$$

This expression seems highly intuitive and easy to accept even without the formal application of the virial theorem. (Every individual term in (4) has its counterpart in eq.(8), section III of Ref.3, derived through a different form of statistical mechanics.)

In (4) the first term on the right is the kinetic pressure of the solvent molecules and the second the contribution of forces acting between the molecules in solvent pairs (*w* for water avoids confusion with *s* for solute, this is illustrated in the left hand panel of Fig.2). The third term $P_{ws}^R$ is the contribution to the pressure of the forces acting between a solvent (water) molecule and a molecule of solute (right hand panel of Fig.2). It is this interaction term that couples the solvent to the solute and can be associated with neither individually. The fourth term is the kinetic pressure of the solute molecules and $P_{ss}^R$ is the contribution to the overall pressure of solute molecules interacting with other solute molecules. For a dilute solution this term can be ignored.

In the left hand compartment of Fig.1 we suppose pure water, the pressure of which is given by

$$p^L = K_w^L + P_{ww}^L \qquad (5)$$

The pressure $p^L$ of the pure water might be 1 atmosphere, yet the kinetic pressure alone in liquid water over 1000 atmospheres. This is largely cancelled by the tension from inter-molecular forces.

When these two compartments are brought into contact through a semi-permeable membrane, water is free to flow from one to the other. The forces driving fluid flow derive from a pressure gradient and so the flow will only be arrested when the pressures of the water on each side are equal. The key to a general understanding of osmotic pressure is the correct identification of those terms on the right in (4) deriving from forces tending to drive water from the right hand (solution) compartment in Fig.1 and so resisting the water pressure in the left hand compartment. They are simply those with a $w$ subscript. These terms exert pressure on a piston but cannot exert pressure on a semi-permeable membrane. In equilibrium, the water can exert no forces on the membrane and consequently $K_w^R$ and $P_{ww}^R$ must be balanced by the water pressure acting on the left hand piston, $p^L$. The term $P_{ws}^R$ is contributed by the forces between water and solute molecules (Fig.2). Because the water can move freely through the membrane and between compartments, the forces between water and solute molecules can exert no pressure on the semi-permeable membrane- this is perhaps less obvious and is discussed in detail in sections IV and V below. Given this property of semi-permeable membranes, common to the classic treatments[5,8], the pressure component $P_{ws}^R$ exerted on the right hand piston, must also be balanced by water pressure on the left. Thus the water pressure in the solution compartment, which in equilibrium must equal the pressure in the pure water, is given by

$$p_w^R = K_w^R + P_{ww}^R + P_{ws}^R \tag{6}$$

The pressure in (6) is perfectly well defined but does not, by itself, have physical significance in the absence of the semi-permeable membrane. It cannot be called a partial pressure in the conventional sense and cannot be expressed purely in terms of the properties of water molecules. (A new descriptor is needed – it would be appropriate to call $p_w^R$ the water (or solvent) osmotic pressure, but at the risk of some confusion. Perhaps *exosmotic pressure* would serve.)

The osmotic pressure is now easily extracted. The condition for no net flow of water from the left hand compartment (Fig.1) into the solution being that the water pressures balance, $p^L = p_w^R$, substitution of (6) into (4) yields

$$p^R = p^L + K_s^R + P_{ss}^R \tag{7}$$

Thus the osmotic pressure is given by

$$\Pi = K_s^R + P_{ss}^R \tag{8}$$

and it does not involve the interaction term[5,8]. This is important because $P_{ws}^R$ is likely, for a liquid solution, to be comparable in magnitude (and, for cohesive forces, of opposite sign) to the solute kinetic pressure, the first term on the right of (8). The second term is the pressure from solute-solute molecular interactions and for a dilute solution can be ignored. Equation (8) is then dominated by the kinetic pressure of the solute (which is why van't Hoff's law looks like an ideal gas relation) but can only be identified as the partial pressure of the solute for the case of dilute gas mixtures. More generally, the second term in (8) cannot be ignored.

Equation (6) is the crux of this paper and both (6) and (8) depend on the cross term $P_{ws}^R$ balanced by pressure in the pure water compartment. This in turn depends on water mobility through the membrane resulting in no pressure on the membrane from this term. That this is so is established through an analysis based on insights contained in the kinetic theory of Ehrenfest[5], discussed in section IV below. It is also intrinsic to the very different kinetic treatment of Joos[8] (section V).

The term $P_{ws}^R$ in (4) usually cannot be split into solute and solvent components but, at least in principle, it can thrust both ways. Should there be a semi-permeable membrane passing the solute (perhaps alcohol) but the water going nowhere, the solute would be driven by an *exosmotic* pressure (cf (6))

$$p_s^R = K_s^R + P_{ss}^R + P_{ws}^R \tag{9}$$

For a dilute (liquid) solution, the left hand side of (9), $p_s^R$, is very much less than the kinetic pressure and the last term on the right hand side is largely responsible. If the solution is not dilute, the question of which is the solute and which the solvent becomes artificial – consider a mixture of water and alcohol. The real distinction is which component is contained by the membrane and which passes freely, exerting no pressure on the membrane; the same cross term appears in both (6) and in (9).

**IV. Ehrenfest's model re-visited**

The treatment given by Ehrenfest[5] is rather different, yet has proved to be highly relevant. In my use of the virial theorem I considered closed containers of on the one hand water and the other of solution, requiring the water pressures to be equal for equilibrium when connected through a semi-permeable membrane. Ehrenfest[5] desired to calculate directly the force exerted by the solute on a semi-permeable membrane. He considered a closed membrane – a bladder – containing solution and immersed in water. He then applied the virial theorem to molecules of solute only, water free to flow through the bladder. He argued that because of uniformity in the interior of the bladder, solute molecules deep inside would experience no net force. Molecules close to the surface could, but he further argued that a molecule of solute close to a pore would be pulled one way by water inside the bladder and the other way by a mirror image water molecule on the far side. For a dilute solution a solute molecule close to the membrane is surrounded by water much as one deep inside. Consequently, solute close to the surface experiences, at least to a very good approximation, no force from water molecules. Ehrenfest envisaged any forces acting on a solute molecule near the membrane as being proportional to the difference in bulk concentration of water on each side. He argued that the pressure exerted by solute-water forces is second order in solute concentration, whereas the kinetic pressure of the solvent is first order. In its original form his argument already validates (6) for dilute solutions.

The idea that in equilibrium the local concentration of water molecules in the region of the semi-permeable membrane is different from the concentration outside the membrane seems mistaken, given that the membrane essentially does not exist for water molecules. Ehrenfest's model is improved by noting that footloose water molecules will move around until there is a common water pressure on both sides of the bladder membrane. Solute molecules close to the membrane will experience no force from the water and water-solute forces contribute no pressure on the interior of Ehrenfest's bladder. (A cartoon image may be helpful here. Think of the membrane as represented by a net in the water, of too fine a mesh to permit passage of neutral buoyancy spheres confined to one side (the solution side) of the net. There is no force exerted by the water on the spheres, nor on the net. The bulk concentration of water molecules on the solution side is nonetheless below the concentration outside, because of the volume excluded by the spheres.)

The results of this refinement of Ehrenfest's model are, first, that there is no requirement to neglect the solute-solute interactions in calculating the virial for the solute molecules; the model then immediately yields (8). Secondly, with solute-water interactions exerting in equilibrium no pressure on the semi-permeable membrane, (6) is validated without any restriction to dilute solutions.

**V. Joos' membrane model**

A different approach to the origin of osmosis treats explicitly forces exerted directly on the solute by the semi-permeable membrane. This treatment apparently originated in the early 1930s, appearing in Joos' book on Theoretical Physics[8] ; a more recent discussion in a broader context can be found in Ref.9. A solute molecule a distance $x$ from the membrane experiences a force $f = -du(x)/dx$, where $u(x)$ is a potential energy term that reaches infinity at $x=0$ and tends rapidly to zero as $x$ increases. For an isothermal solution the number density of solute molecules is given as a function of $x$ by the Boltzmann factor

$$n(x) = n_\infty \exp(-u(x)/kT) \quad (10)$$

The force acting in unit volume of solution is given by

$$f(x) = -n(x) du(x)/dx \quad (11)$$

At equilibrium there is no bulk motion of fluid and this force must be balanced by a pressure gradient in the solution, the pressure rising with increasing $x$ (just as the gravitational force on an element of water is opposed by pressure increasing with depth). Thus

$$\frac{dp(x)}{dx} = -n(x)\frac{du(x)}{dx} \quad (12)$$

Differentiating (10)

$$\frac{dn(x)}{dx} = -\frac{1}{kT} n(x) \frac{du(x)}{dx} \quad (13)$$

Then substituting (13) into (12),

$$\frac{dp(x)}{dx} = kT \frac{dn(x)}{dx} \qquad (14)$$

Equation (14) integrates to yield *p(x)*

$$p(x) - p(0) = kT(n(x) - n(0)) \qquad (15)$$

The boundary condition is $n(0) = 0$, because $x = 0$ defines the plane never reached by solute molecules invading the repulsive potential. The pressure $p(0)$ is the solution pressure at the plane $x = 0$ and this is in pure water. Thus $p(0)$ is the pressure acting to drive water from solution and out through the membrane and for equilibrium this must equal the water pressure on the pure water side. Water pressure is, in equilibrium, uniform from the water side, through the force field representing the membrane and on into the solution. Joos' semi-permeable membrane model has the explicit feature that water exerts no force on the solute molecules and hence that water-solute interactions exert no force on the membrane. It is important that the relation (15) does not depend at all on the form of the repulsive potential $u(x)$, other than that it becomes effectively infinite at $x = 0$ and falls very rapidly to zero for positive *x* - the limit is a step function and in this limit the model membrane becomes the solute reflecting wall assumed in the treatments of section III and in Ref. 5.

Once clear of the force field established by the membrane, the pressure in the solution is

$$p = p_0 + kTn_\infty \qquad (16)$$

where $n_\infty$ is the number density of the solute molecules in the solution clear of the force field, *p* is the pressure $p^R$ of (7) above and $p(0)$ the pressure $p^L$. The law of van't Hoff follows immediately.

This ingenious argument is, within the assumptions made, correct. It applies to a static equilibrium in which there is no bulk transport of any kind, and no kind of viscous drag. It equates water pressures across the semi-permeable membrane, just as in the treatment of section III. It differs in how the water pressure in solution is calculated. The osmotic pressure is simply $\Pi = kTn_\infty$, the kinetic pressure only. It does not contain a term from solute-solute interaction (because (1) is only valid for a dilute solution) and it does not contain a term from solute-solvent interaction, for the reasons already discussed. This is intrinsic to Joos' treatment and in accord with (6) above.

**VI. Discussion**

The idea that osmosis is driven by the difference in water pressures across the semi-permeable membrane apparently originated with G. Hulett in 1903. It was taken up by H. T. Hammel and collaborators [10], see also Ref.11 . That formulation was heavily criticised [2,11] ; in Ref. 3 the authors express their opinion that the Hammel solvent tension theory is discredited by the discussion in Ref.11. It would seem that the very notion of osmosis being driven by a difference in water pressures acquired a taint that yet persists. However, a flawed formulation does not necessarily invalidate the underlying ideas. (In Ref.11 the membrane is modelled in such a way that water is driven through it by a pressure gradient in pure water within the membrane itself, that water pressure reaching the value that would be predicted by (6) very close to the solution face of the membrane.)

The analysis of sections II and III above, in terms of the virial theorem, is algebraically complete and does not depend on any modelling of the semi-permeable membrane. All that is required is that solvent and solution are fluids in thermal equilibrium, separated by a membrane that is semi-permeable. Equations (4) and (5) are exact for particles confined within some specified volume; equations (7) and (8) then provide a complete description of osmosis and osmotic pressure. The process is driven by the difference in water pressure between the pure solvent and solution sides. On the solution side the water pressure cannot in general be regarded as a partial pressure; part of the reason for this is that the term $P_{ws}^R$ in (6) couples the solvent to the solute. The water pressure (6) is nonetheless well defined and equating (5) and (6) yields the condition for osmotic pressure in agreement with the result from the entirely different culture of classical thermodynamics (in which the role of entropy of mixing is to translate into thermodynamics the fact that there is less water in a volume of solution than in the same volume of pure water).

The virial theorem analysis also casts light on treatments deriving the pressure difference required to stop solvent flow by equating chemical potentials. In Ref.3 the authors write down a partition function for a general two fluid mixture and calculate the Helmholtz free energy; thence both chemical potentials and pressures. The general expression for pressure contains terms precisely corresponding to those in (6) above. Equating the water chemical potentials between pure water and solution yields van't Hoff's law. It may not be generally appreciated that equating my water pressure terms extracted from eq.(8) of Ref.3 yields exactly the same conditions, but this is easily demonstrated.

The derivation from the virial theorem also applies equally well to mixtures of gases as to liquid solutions. While equating (5) and (6) looks very much like equating partial pressures, this is not an acceptable analogy, even in the event of the coupling term $P_{ws}^R$ being negligible. The realistic example of this condition is of course mixtures of gases, where all intermolecular interactions can be ignored at low pressures. The phenomenon of osmosis exists independent of the nature of the coupling term $P_{ws}^R$, which does not appear in the expression (8) for the osmotic pressure. It does not matter whether $P_{ws}^R$ is positive (repulsive forces), negative (attractive forces) or negligible and it is useful to consider the hypothetical case of negligible interaction between solute and liquid solvent. The two are then decoupled and can be considered independently. The solute is confined by the semi-permeable membrane and the water pressure terms must be equal on the two sides. The osmotic pressure is still given by (8). (Joos' treatment[8] is still valid in this hypothetical case.) Thus the role of the semi-permeable membrane in osmosis is to keep the solute in the right compartment and to hold the resulting pressure difference between the two compartments.

## VII. CONCLUSIONS

Application of the virial theorem in classical mechanics to osmosis shows rigorously that the equilibrium condition corresponds to equalization of solvent pressures on the two sides of a semi-permeable membrane. Osmotic pressure so calculated from classical mechanics agrees with the result from classical thermodynamics. The calculation of the osmotic pressure is in both cases static, from equilibrium conditions: in thermodynamics the chemical potentials of water are equated across the semi-permeable membrane; when using the virial theorem it is the water pressures that are equated. This analysis has illuminated the statistical mechanics of Ref.3, which can employ either route, linking classical thermodynamics to classical mechanics. It has also cast light upon and agrees with the rather different kinetic treatments of Ref.5 and Ref. 8, which in turn illuminate a property of semi-permeable membranes that determines the composition of water pressure in solution.

Although all the above treatments correspond to equating pressures at equilibrium, it is clear that slow changes can be treated through eqs. (6)-(8). A slight increase in the pressure $p^L$ produces a pressure gradient driving solvent into the solution compartment. Keep pushing on the left hand piston and solvent is steadily driven into the solution. However, if an osmotic system relaxes in a way that is not quasi-static, the details will be hydro-dynamical rather than hydrostatic and viscous forces could well enter, but the system is yet driven by differences in water pressure. Water 'finding its own level' is an archaic way of putting it. For the static and quasi-static cases, qualitative descriptions in terms of solute recoiling from the membrane and transferring momentum to solute seem to have no place.

This treatment, using classical mechanics and the virial theorem, addresses the mechanism by which the spectacular osmotic pressures are generated and maintained, illuminating the dynamics of osmosis and the process of reverse osmosis: an osmotic pressure of one atmosphere represents an unbalance of one part in $10^3$ between kinetic and cohesive pressures. The virial theorem thus draws together many different strands[3,5,8] into a unified picture of osmosis driven by differential water pressures.

## ACKNOWLEDGEMENTS

Borg's [4] achievement was to have realized how the virial theorem in classical mechanics could be profitably applied to osmosis. His stimulating work made this paper possible.
I thank D R Bowler for useful correspondence and J D P Bowler for turning sketches into figures.

[a] e-mail: michael.bowler@physics.ox.ac.uk